\documentclass[twocolumn,twoside,journal]{IEEEtran}

\usepackage{algorithm}
\usepackage{balance}

\usepackage{cite}

\ifCLASSINFOpdf
 \usepackage[pdftex]{graphicx}
 \graphicspath{{../pdf/}{../jpeg/}}
\else
 \usepackage[dvips]{graphicx}
 \graphicspath{{./eps/}}
 \DeclareGraphicsExtensions{.eps}
\fi
\usepackage{amsmath}
\usepackage{algorithmic}

\usepackage{array}
\usepackage{url}

\usepackage{amsfonts}
\usepackage{mathrsfs}
\usepackage{latexsym}
\usepackage{subfigure}
\usepackage{mathtools}

\usepackage{bm}
\usepackage{amssymb}
\usepackage{amsthm}

\usepackage{multirow}
\usepackage{arydshln}

\usepackage{comment}

\usepackage{balance}

\newtheorem{theorem}{Theorem}

\newtheorem{remark}{Remark}

\usepackage{enumitem}
\setlist[itemize]{align=parleft,left=0pt..1em}

\begin{document}
%
\title{A Persistent-Excitation-Free Method for System Disturbance Estimation Using Concurrent Learning}
%
%
%

\author{
Zengjie~Zhang,~\IEEEmembership{Member,~IEEE,}
 Fangzhou~Liu$^*$,~\IEEEmembership{Member,~IEEE,}
 Tong~Liu,\\
 Jianbin~Qiu,~\IEEEmembership{Senior Member,~IEEE,}
 Martin Buss,~\IEEEmembership{Fellow,~IEEE,}

\thanks{This work was supported by the National Natural Science Foundation of China (U21B6001, 62273121, and 62173147).}
\thanks{* Corresponding author.}
\thanks{Z. Zhang is with the Department of Electrical Engineering, Eindhoven University of Technology, 5600 MB Eindhoven, Netherlands (e-mail: z.zhang3@tue.nl).
}
\thanks{F. Liu and J. Qiu are with the Research Institute of Intelligent Control and Systems, Harbin Institute of Technology, 150001 Harbin, China (email: fangzhou.liu; jbqiu@hit.edu.cn).}

\thanks{T. Liu and M. Buss are with the Chair of Automatic Control Engineering, Technical University of Munich, 80333 Munich, Germany (e-mail: tong.liu; mb@tum.de).}
}

\maketitle


\begin{abstract}
Observer-based methods are widely used to estimate the disturbances of different dynamic systems. However, a drawback of the conventional disturbance observers is that they all assume persistent excitation (PE) of the systems. As a result, they may lead to poor estimation precision when PE is not ensured, for instance, when the disturbance gain of the system is close to the singularity. In this paper, we propose a novel disturbance observer based on concurrent learning (CL) with time-variant history stacks, which ensures high estimation precision even in PE-free cases. The disturbance observer is designed in both continuous and discrete time. The estimation errors of the proposed method are proved to converge to a bounded set using the Lyapunov method. A history-sample-selection procedure is proposed to reduce the estimation error caused by the accumulation of old history samples. A simulation study on epidemic control shows that the proposed method produces higher estimation precision than the conventional disturbance observer when PE is not satisfied. This justifies the correctness of the proposed CL-based disturbance observer and verifies its applicability to solving practical problems.
\end{abstract}


%
\IEEEpeerreviewmaketitle

\section{Introduction}
%
%
%
%


\IEEEPARstart{T}{he} {disturbance} is an important reason for performance degradation of many practical systems, such as switched systems~\cite{du2020semi}, circuit systems~\cite{wang2020extended}, and multi-agent systems~\cite{cheng2021fixed}. Disturbances are often recognized as unexpected unknown inputs of the systems, such as actuator faults~\cite{chen2022fault}, external impacts~\cite{zhang2020online}, impulses~\cite{tang2020input}, or vibrations~\cite{zhao2019vibration}. To attenuate and mitigate the influences of the disturbances on system performance, various disturbance-tolerant control methods are proposed to compensate for the disturbance effects by refining the control inputs~\cite{zhang2020safe, zhang2020robust, wang2019continuous}.
These methods require precise estimation of the system disturbances. The most effective methods for disturbance estimation are mainly based on analytical redundancy technology, namely disturbance observers. A classical type of disturbance observer is the unknown-input observer that reconstructs the disturbances using the linear observer theory. Nevertheless, the main drawback is that it assumes the smoothness of the system nonlinearity~\cite{chakrabarty2018state}. Also, this method requires the decoupling between the disturbance and the observation of the system. To improve the robustness of the estimation, the sliding mode observer is proposed to compensate for the unmodeled dynamics using high-frequency switching~\cite{farahani2022sliding}. Besides, the nonlinear disturbance observer\cite{aljuboury2022robust, liu2022nonlinear} utilizes feedback linearization to construct linear error dynamics and provide precise disturbance estimation. It does not require the decoupling property but needs the derivatives of the system states~\cite{zhao2019boundary, DeLuca2003}. To obtain the exact state derivatives, the second-order and the integral sliding mode observers~\cite{gu2022disturbance, zhang2019integral} are proposed. These are the most representative disturbance-estimation methods presented in the previous work. Recent surveys on the variants of these methods can be referred to in~\cite{chen2015disturbance, sariyildiz2019disturbance}. 

Nevertheless, the conventional observer-based methods are only applicable to cases where the persistent excitation (PE) condition is satisfied for the system. PE is an important concept in system identification to address that the system is sufficiently actuated by a rich amount of spectral components of the input signals~\cite{lu2016improved}. It is a necessary condition to guarantee that the system parameter or structure to be identified can be precisely reconstructed under the actuation of these input signals. If the input signals of a system do not satisfy the PE condition, there may exist large deviations between the identified system parameters and their true values~\cite{liu2021adaptive}. Many works are devoted to solving system identification problems without PE conditions~\cite{kersting2019recursive}. Nevertheless, disturbance estimation under PE-free conditions has not attracted much attention. To the best knowledge of the authors, there has not been a work that solves the estimation of disturbance or time-variant parameters for PE-free systems. The main reason is that PE-free cases are not very common in practice. Most of the practical systems support the PE assumptions since they usually have non-singular disturbance gains~\cite{chen2015disturbance}. For example, the disturbance gain is a constant non-singular matrix in~\cite{haes2019wams}. Also, the disturbance gain of a robotic system is usually its inverse inertia which is typically always positive-definite~\cite{zhang2019integral, zhang2022disturbance}. The conventional disturbance observers have no problems when applied to these systems. However, there exist some systems of which the disturbance gain may become singular in some states, making the systems lose PE. Examples of such systems include the networked epidemic model~\cite{liu2019robust}, the population dynamic model~\cite{song1988population}, the underactuated robot model with external collisions~\cite{katsura2002wheelchair}, circuit network with noise~\cite{uchida2019incentivizing}, or general networked systems with impulse disturbances~\cite{tang2022event}, which will be detailedly elaborated in Sec.~\ref{sec:preC}. For these systems, the conventional disturbance-observer-based methods may produce large estimation errors when the systems are close to the singularity states. This issue attracted our attention due to our previous work on the control and filtering of networked epidemic models~\cite{liu2019robust, liu2021optimal}. We believe that investigating PE-free disturbance estimation is valuable work, considering the completeness of the observation theory. This work is the first attempt to solve this problem.

The system disturbance losing PE reflects the lack of global diffeomorphic mappings for the disturbance-output feedback linearization of the system, which will be explained in Sec.~\ref{sec:preA} and Sec.~\ref{sec:preB}. An effective method for PE-free estimation problems is concurrent learning (CL). Since proposed in~\cite{chowdhary2010concurrent}, CL is widely applied to system identification~\cite{
walters2018online,du2021online}, adaptive control~\cite{
zhang2022integrated}, robust control~\cite{
zhang2017event}, optimal control~\cite{
kamalapurkar2016efficient}, observer design~\cite{yue2022data}, and differential games~\cite{kamalapurkar2014concurrent}. By utilizing the history stack, a queue structure storing history system states and inputs, CL ensures precise approximation of system parameters without PE~\cite{chowdhary2013concurrenta
}. However, compared to constant parameters, estimating time-variant disturbances is challenging due to the accumulated errors brought up by the history stacks. Even though CL is applied to state observation~\cite{rotithor2020full} where the accumulated errors are avoided by utilizing the known intrinsic dynamics, similar techniques can not be applied to disturbances that are exerted by unknown extrinsic dynamics. Another solution to precisely estimate disturbance for PE-free systems is to utilize the higher-order derivatives of the system states to reconstruct the disturbance~\cite{Chen2000, DeLuca2003}, which, however, increases the complexity of the observers. Also, solving state derivatives is difficult in practice due to the existence of noise. Thus, such methods are not widely used by the previous work due to lack of applicability.

In this paper, we propose a novel CL-based disturbance observer to estimate the disturbance of PE-free systems. The systems we are concerned with have state-dependent disturbance gains which become or get close to singularity under some states. The main contribution of this work is reflected from the following two perspectives. 
\begin{itemize}
\item Firstly, for the first time, we present how to use CL to precisely estimate disturbance for PE-free systems. Specifically, an extrinsic model adapted from the previous work~\cite{Chen2000} is used to approximate the external dynamics of the disturbance. Two time-variant history stacks are constructed to refine the updates of the disturbance estimation. Also, we define and analyze the accumulated errors brought up by CL, which is unique and nontrivial for disturbance estimation problems. The Lyapunov-based method is used to address the boundedness of the estimation error. \item Secondly, we propose a history-sample-selection procedure to reduce the accumulated errors caused by CL. We apply the proposed CL-based disturbance observer to a networked epidemic model which is a very practical model for epidemics prediction and reaction. Specifically, the proposed observer is used to predict the infection rates of a simulated epidemic process. The precise prediction of the infection rates indicates the success of the proposed method. 
\end{itemize}
The simulation results indicate that the proposed methods can be realized on a normal PC without graphics cards. Compared to the conventional disturbance observers, the proposed CL-based disturbance observer has lower efficiency due to the stacking of the history data. Nevertheless, it wins with higher estimation precision. Our method is promising to promote the precision of unknown input estimation for large-scale network systems such as circuit networks.

The rest of this paper is organized as follows, section~\ref{sec:pre} formulates the disturbance estimation problem, and section~\ref{sec:main} presents the main results of the CL-based disturbance observer. In section~\ref{sec:simu}, an epidemic-control case is simulated to validate the feasibility and efficacy of the proposed method. Finally, section~\ref{sec:concld} concludes the paper.

\textit{Notations:} $\mathbb{R}_{\geq 0}$ and $\mathbb{R}^+$ are the set of non-negative and positive real numbers. $\mathbb{N}$, $\mathbb{N}_{\geq 0}$, and $\mathbb{N}^+$ are the sets of integers, non-negative integers, and positive integers, respectively. For a vector $x \!\in\! \mathbb{R}^n$, $x_i$ or $(x)_i$ denotes its $i$-th element, $i = 1,\cdots,n$, $\|x\|$ is its 2-norm, and $\mathrm{diag}(x) \in \mathbb{R}^{n \times n}$ is the diagonal matrix composed of $x$. For any differentiable vector function $v(x)\!:\! \mathbb{R}^n \! \rightarrow \! \mathbb{R}^m$, $n,m \!\in\! \mathbb{N}^+$, $\nabla v(x) \!=\! {\partial v(x)}/{\partial x} \!\in\! \mathbb{R}^{m \times n}$ denotes its gradient. For any $h(x): \mathbb{R}^n \rightarrow \mathbb{R}$ and $f(x):\mathbb{R}^n \rightarrow \mathbb{R}^n$, $L_fh(x) \!=\! \nabla h(x) f$ is the Lie-derivative of $h$ for $f$ and $L_f^mh(x) \!=\! \underbrace{L_fL_f\cdots L_f}_{m}h(x)$ is the $m$-th order Lie-derivative. For a matrix $M \!\in\! \mathbb{R}^{m \times n}$, $\|M\|$ denotes its spectral radius. $I$ and $O$ are the identity and zero matrices.

\section{Preliminaries}\label{sec:pre}

In this section, we present the preliminary knowledge that is needed to interpret the results of our work. Firstly, we formulate the disturbance estimation problem for a nonlinear system. Secondly, we address the PE-free conditions for the introduced system and explain the main challenges of PE-free disturbance estimation. Finally, we introduce three dynamic models as examples that can meet PE-free conditions.

\subsection{Problem Formulation}\label{sec:preA}

We consider the following general nonlinear system,
\begin{equation}\label{eq:sysori}
 \dot{x}(t) = f(x(t)) + E(x(t))u(t) + G(x(t))d(t),
\end{equation}
where $x(t) \in \Omega_x \subseteq \mathbb{R}^n$ is the system state, $\Omega_x$ is the feasible state domain, $u(t) \in \mathbb{R}^m$ and $d(t) \in \mathbb{R}^p$ are respectively the control input and time-dependent disturbance of the system, $f\!:\! \mathbb{R}^n \!\rightarrow\! \mathbb{R}^n$, $E \!:\! \mathbb{R}^n \!\rightarrow\! \mathbb{R}^{n \times m}$, and $G \!:\! \mathbb{R}^n \!\rightarrow\! \mathbb{R}^{n \times p}$ are smooth vector functions. The problem studied in this paper is to estimate the disturbance $d(t)$ using the measurable state $x(t)$ and its history data $x(t_i)$, $0 \!<\! t_1 \!<\! t_2 \!<\! \cdots \!<\! t_i \!<\!\cdots \!<\! t$. To clarify the PE-free situation for system~\eqref{eq:sysori}, we reformulate it into a regularized form using feedback linearization. Suppose the existence of a smooth vector function $h(x) \in \mathbb{R}^p$ and $p$ integer scalars $r_1$, $\cdots$, $r_p$ $\in \mathbb{N}^+$, such that the following conditions hold.
\begin{enumerate}
\item \label{it:m1} There exists $r \in \mathbb{N}^+$, such that $p \leq r = \sum^p_{i=1} r_i \leq n$.
\item \label{it:m2} For all $x \in \Omega_x$, $i = 1$, $\cdots$, $p$, and $k = 1$, $\cdots$, $r_i-1$, 
\begin{equation*}
[\, L_{g_1} L^{k-1}_f h_i(x)~ ~ \cdots ~ L_{g_p} L^{k-1}_f h_i(x) \,] = 0
\end{equation*}
where $g_j$, $j = 1$, $\cdots$, $p$, is the $j$-th column of $G(x)$.
\item \label{it:m3} There exists $x \in \Omega_x$, such that for all $i = 1$, $\cdots$, $p$, 
\begin{equation*}
l_i(x) = [\, L_{g_1} L^{r_i-1}_f ~ \cdots ~ L_{g_p} L^{r_i-1}_f h_i(x) \,] \neq 0.
\end{equation*}
\end{enumerate}
Then, for $x \in \Omega_x$ where \ref{it:m3}) holds, there exist diffeomorphic mappings 
$z \!=\! \psi(x) \!=\![\,
  \psi_1(x) ~
  \cdots ~
  \psi_{r}(x) \, ]^{\top\!}  \in \mathbb{R}^{r}$ and $w \!=\! \phi(x) \!=\! [ \,
  \phi_1(x) ~
  \cdots ~
  \phi_{n-r}(x) \,
 ]^{\top\!} \!\in\! \mathbb{R}^{n-r}$, 
 where for each $i = 1, \cdots, r$, $\psi_i(x) = [ \,
  h_i(x) ~
  L_f h_i(x) ~
  \cdots ~
  L^{r_i-1}_f h_i(x) \,
 ]$, and the elements of $\phi$ satisfy $L_{g_j}\phi_i(x) = 0$, for $i = 1,\cdots, n- r$, $j= 1,\cdots, p$, such that system (\ref{eq:sysori}) can be represented as
\begin{subequations}\label{eq:sysreg}
\begin{alignat}{2}
 \dot{w} &= \eta(x,u) \\
 \dot{z} &= \gamma(x,u) + B L(x) d(t), \label{eq:sysreg2} \\
 y & = Cz,
\end{alignat}
\end{subequations}
where $\gamma(x,u)=A \psi(x) + \iota(x,u) + B \alpha(x)$, and $A \in \mathbb{R}^{r \times r}$, $B \in \mathbb{R}^{r \times p}$, $C \in \mathbb{R}^{p \times r}$, $\alpha(x) \in \mathbb{R}^p$ and $L(x) \in \mathbb{R}^{p \times p}$ are
\begin{equation}\label{eq:syspar1}
\begin{split}
&A = \mathrm{diag} \left( 
  A_1, \,
  A_2, \,
  \cdots, \, 
  A_p \right),\\
 &B = \left[\, B^{\top\!}_1 ~ B^{\top\!}_2~ \cdots ~ B^{\top\!}_p \, \right]^{\top\!}, ~
 C = \left[ \, C_1 ~ C_2 ~ \cdots ~ C_p\, \right], \\
 &\alpha(x) = [ \,
  L_f^{r_1} h_1(x) ~ L_f^{r_2} h_2(x) ~ \cdots ~ L_f^{r_p} h_p(x)
 \, ]^{\top\!}, \\
 &L(x) = [ \, l_1^{\top\!}(x)~l_2^{\top\!}(x)~\cdots~l_p^{\top\!}(x) \, ]^{\top},
\end{split}
\end{equation}
where $A_i \in \mathbb{R}^{r_i \times r_i}$, $B_i \in \mathbb{R}^{r_i \times p}$ and $C_i \in \mathbb{R}^{p \times r_i}$, $i = 1,\cdots,p$ are sub-blocks of matrices $A$, $B$ and $C$, 
\begin{equation}\label{eq:syspar2}
A_i \!=\!\! \left[\!\! \begin{array}{cc}
  &\!\!I_{(r_i-1)} \\ 
  O_{1 \times 1}\!\!\!\!& 
 \end{array} \!\!\right]\!, \,
B_i \!=\!\! \left[\!\! \begin{array}{cc}
  &\!\!O_{(i-1) \times (r_i-1)} \\ 
  I_{1 \times 1}\!\!& \\ 
  &\!\!O_{(p-i) \times (r_i-1)}
 \end{array}\!\! \right]^{\!\top\!}\!,
\end{equation}
and $C_i \!=\! B^{\top\!}_i$, where all blank positions in the matrices are zero. The state-input-dependent smooth functions $\iota(x,u) \in \mathbb{R}^{r}$ and $\eta(x,u) \in \mathbb{R}^{n-r}$ are determined by the mappings $\psi$ and $\phi$. Specifically, the $i$-th elements of $\iota(x,u)$ and $\eta(x,u)$, $i=1,2,\cdots,n-r$, are respectively represented as
\begin{equation*}
\begin{split}
 \iota_i(x,u) =& \textstyle \sum^{p}_{j=1} \sum^{r_i}_{k=1} L_{e_j}L_f^{r_i-k}h_i(x)u^{(k-1)}_j(t),\\
 \eta_i(x,u) =& \nabla \phi_i(x) \left( f(x) + E(x)u \right),
\end{split}
\end{equation*}
where $e_j$ is the $j$-th column vector of $E(x)$ and $u^{(k-1)}_j(t)$ is the $j$-th element of $u^{(k-1)}(t)$, the $(k\!-\!1)$-th derivative of $u(t)$. Note that we assume $u(t)$ is $\max_i(r_i)-1$-times differentiable. The matrix $L(x)$ in (\ref{eq:sysreg}) is the state-dependent disturbance gain.

\subsection{The PE-Free Conditions}\label{sec:preB}

For the regularized system (\ref{eq:sysreg}), if the diffeomorphic mappings $\psi(x)$ and $\phi(x)$ globally exist for all $x \in \Omega_x$, the regularized form (\ref{eq:sysreg}) holds in the entire state domain $\Omega_x$~\cite{khalil2002nonlinear} and the disturbance gain $L(x)$ contains no zero rows. If $L(x)$ is further assumed to be non-singular for all $x \in \Omega_x$ (such as the Euler-Lagrangian systems~\cite{zhang2019integral}), for any $t \in \mathbb{R}_{\geq 0}$, there exist $T, \bar{\tau} \in \mathbb{R}^+$, such that $d(t)$ and $L(x)$ satisfies
\begin{equation}\label{eq:pers}
\textstyle \int^{t+T}_t L(x(\tau))d(\tau)\!\left(L(x(\tau)) d(\tau) \right)^{\!\top\!} \mathrm{d}\tau \geq \bar{\tau} I,
\end{equation} 
which indicates that $d(t)$ persistently excites the system~\cite{chowdhary2010concurrent}. The conventional disturbance observers are eligible for this situation. However, for a certain system for which no global diffeomorphic mappings exist, condition (c) in Sec~\ref{sec:pre} does not hold for all $x \in \Omega_x$. Instead, there exists $x \in \Omega_x$ such that some rows of $L(x)$ become zeros and $L(x)$ becomes singular. In this situation, PE is not ensured and the conventional PE-based observers can provide imprecise estimation results. A typical example is the network epidemic model which will be introduced in the simulation study in section~\ref{sec:simu}, where we will see that the conventional disturbance observer produces large estimation errors in a PE-free situation.
Besides, the mobile robots with external forces in singular positions~\cite{katsura2002wheelchair} and the population system with empty regions~\cite{song1988population} also address the similar issue.
The critical point in resolving this issue is to exploit the history system data in the closed loop of the observation rather than only the system state.
To solve this problem, in this paper, we design a CL-based disturbance observer by exploiting the history data of the system, such that precise estimation is ensured even when PE is not satisfied.

\subsection{Examples of Disturbance-PE-Free Systems}\label{sec:preC}

PE-free conditions are not typical for practical system disturbances. This is also why PE-free disturbance estimation has not attracted much attention. Nevertheless, there exist some systems for which PE-free cases should be incorporated. Here, we raise three examples.

\subsubsection{Networked epidemic model\,\cite{liu2019robust}} The epidemic process over a social network with $n \in \mathbb{N}^+$ nodes is represented by the following continuous-time model
\begin{equation}\label{eq:contep}
\dot{x}(t) = (I- \mathrm{diag}(x(t)))W \mathrm{diag}(x(t)) d(t) - \mathrm{diag}(x(t)) \delta(t),
\end{equation}
where $x(t) \in \mathbb{R}^n$ is the infection probabilities of the nodes, $W \in \mathbb{R}^{n \times n}$ is the adjacency matrix of the network, and $d(t), \delta(t) \in \mathbb{R}^n$ are respectively the infection and curing rates of the social nodes. This model will be interpreted in detail in Sec.~\ref{sec:chal}. Recognizing the infection rate $d(t)$ as the unknown disturbance of the system, this model is in a regular form as \eqref{eq:sysreg} without performing feedback linearization. In this sense, $(I- \mathrm{diag}(x(t)))W \mathrm{diag}(x(t))$ is the disturbance gain, and the system loses PE when at least one individual has zero or one infection probability.

\subsubsection{Population model\,\cite{song1988population}} In a certain region that contains $n \in \mathbb{N}^+$ areas, the continuous-time dynamic model of a population system is denoted as $\dot{x}(t) = H x(t) + \mathrm{diag}(x(t)) F b(t) + w(t)$, where $x(t) \in \mathbb{R}^n$ is the population in different areas, $H \in \mathbb{R}^{n \times n}$ is the population transition matrix, $b(t) \in \mathbb{R}^{n}$ is the fertility rate, $F \in \mathbb{R}^{n \times n}$ is the fertility matrix and $w(t)$ is a vector depicting the migration. If the fertility $Fb(t)$ is recognized as a disturbance, the system complies with the regulated form in (\ref{eq:sysreg}). The system loses PE when the disturbance gain $\mathrm{diag}(x_t)$ becomes singular when the population of at least one area decreases to zero. 

\subsubsection{Wheeled robot\,\cite{katsura2002wheelchair}} The continuous-time dynamic model of a unicycle robot with collision force reads $M_{\theta} \ddot{\theta} = \tau + J_{\theta}^{\top\!}F$, where $\theta \in \mathbb{R}^2$ are the rotation angles of the robot wheels, $M_{\theta}, J_{\theta} \in \mathbb{R}^{2 \times 2}$ are the inertia and Jacobian matrices, $\tau \in \mathbb{R}^2$ is the actuation torque, and $F \in \mathbb{R}^2$ is the collision force exerted on the mobile robot. If the collision force $F$ is recognized as a disturbance, the system loses PE when the disturbance gain $m_{\theta}^{-1} J_{\theta}^{\top\!}$ becomes singular. This condition is also referred to as the \textit{singular configuration} in robotics.

For all these systems, the conventional disturbance estimation methods may lead to large errors due to not incorporating PE-free conditions. In Sec.~\ref{sec:simu}, we use a networked epidemic model to address how CL is used to resolve this issue.

\section{Main Results}\label{sec:main}

In this section, we introduce the design of the CL-based PE-free disturbance observer. In subsections \ref{sec:eccd} and \ref{sec:dedist}, we present the continuous-time and discrete-time forms of the observer, respectively. The proofs of the estimation-error convergence are provided using the Lyapunov methods. Then, the accumulated errors caused by the history stacks are analyzed. To restrict the accumulated errors, we present the history-sample-selection procedure in subsection \ref{sec:hissm}.

\subsection{Disturbance Observer in Continuous Time}\label{sec:eccd}

Most of the conventional disturbance estimation methods use predefined disturbance models to provide the necessary prior knowledge of the disturbance. In this paper, we use the following disturbance observer model adapted from~\cite{Chen2000},
\begin{equation}\label{eq:filterob}
\dot{\hat{d}}(t) = \Lambda \hat{d}(t) - \Lambda d(t),
\end{equation}
where $\hat{d}(t) \in \mathbb{R}^p$ is the estimated value of the disturbance $d(t)$ and $\Lambda = \mathrm{diag}\!\left([\,\lambda_1,\, \lambda_2,\, \cdots,\, \lambda_p\,]\right)$ is a constant Hurwitz and diagonal matrix, where $\lambda_i < 0$ for all $i = 1,\cdots,p$. In this sense, (\ref{eq:filterob}) serves as a linear low-pass filter of $d(t)$ or a linear-invariant system with unknown input $d(t)$. The target of the disturbance estimation problem is to precisely estimate $d(t)$ in a real-time manner. Another commonly used disturbance model is the \textit{exogenously-driven model} addressed in~\cite{Chen2004,Jiang2006a,7300416}, which assumes that the dynamic model of the disturbance is precisely depicted by a linear time-invariant system with unknown initial conditions. Actually, the disturbance model~\eqref{eq:filterob} is a generalized version of the \textit{exogenously-driven model} since we do not have strict assumptions on the dynamic model of the disturbance $d(t)$.
Based on this, we present the following CL-based disturbance observer

\begin{equation}\label{eq:obs}
\dot{\hat{d}}(t) = \left( \Lambda \!-\!  \kappa S(\tau_s,t) \right)\! \hat{d}(t) + \kappa X(\tau_s,t),
\end{equation}
where $\kappa \in \mathbb{R}^+$ is a constant gain parameter, $S(\tau_s,t)$ and $X(\tau_s,t)$ are the history stacks defined as,
\begin{equation*}
\textstyle S(\tau_s,t) = \sum^{n_s}_{j=1} S(t_j,t),~
X(\tau_s,t) = \sum^{n_s}_{j=1} X(t_j,t)
\end{equation*}
where $n_s \in \mathbb{N}^+$ is the depth of the stacks, $\tau_s \!=\! \{t_1$, $t_2$, $\cdots$, $t_{n_s}\}$ is a queue that contains the sampling instants of the samples $x(\tau_s)\!=\!\{x(t_1)$, $x(t_2)$, $\cdots$, $x(t_{n_s}) \}$, where we assume the samples are ordered by the sampling sequence, i.e., $0 \!<\! t_1 \!<\! t_2 \!\leq\! \cdots \!\leq\! t_{n_s} \!\leq\! t$. For each $j=1,2,\cdots,n_s$, 
\begin{subequations}\label{eq:st}
\begin{alignat}{2}
\textstyle S(t_j,t) \,&= e^{\Lambda(t_{j}\!- t)} L^{\top}_j L_j e^{\Lambda(t_{j}- t)}, \label{eq:s}\\
\textstyle X(t_j,t) \,&= e^{\Lambda(t_{j}\!- t)} L^{\top}_j B^{\top} \zeta_j, \label{eq:r}
\end{alignat}
\end{subequations}
where $L_j = L(x(t_j))$ and $\zeta_j$ is a difference term
\begin{equation}\label{eq:zeta}
\zeta_j =\gamma(x(t_{j}), u(t_{j})) - \nabla \psi(x(t_j)) \dot{x}(t_{j}),
\end{equation}
where $u(\tau_s)\!=\!\{u(t_{1}), u(t_2), \cdots, u(t_{{n_s}}) \}$, and $\dot{x}(\tau_s)\!=\!\{\dot{x}(t_{1})$, $\dot{x}(t_2)$, $\cdots$, $\dot{x}(t_{{n_s}}) \}$ are the inputs and state-derivatives at the history sampling instants. From (\ref{eq:s}), it is noticed that $S(t_j,t)$ is symmetrically semi-positive definite for all $j=1$, $2$, $\cdots$, $n_s$ since $\Lambda$ is diagonal. Thus, $S(\tau_s,t)$ is also symmetrically semi-positive definite.

\begin{remark}\label{rm:boundelt}
The time-variant history stacks $S(\tau_s,t)$ and $X(\tau_s,t)$ are the critical technical points of the CL-based disturbance observer (\ref{eq:obs}). Different from the conventional CL methods in~\cite{chowdhary2013concurrenta, kersting2014concurrent}, the history stacks in this paper contain a time-variant coefficient $e^{\Lambda(t_j-t)}$ for each sample $x_j$, $t_j \in \tau_s$, where $\Lambda$ comes from the basic observer model (\ref{eq:filterob}) and depicts its filtering bandwidth. The state derivatives $\dot{x}(\tau_s)$ used to construct the stack $X(\tau_s,t)$ can be estimated using exact differentiator~\cite{levant1998robust} or derivative estimator~\cite{kamalapurkar2017concurrent} based methods, which is beyond the scope of this paper. 
\end{remark}

\begin{remark}\label{rm:phyt}
In the disturbance model~(\ref{eq:obs}), the diagonal matrix $\Lambda$ is designed to be Hurwitz, which may lead to the infinity of $e^{\Lambda(t_j-t)}$ as $t-t_j \rightarrow +\infty$. To avoid this, it is kept in mind that an upper limit should be exerted on $t-t_j$ for any sample. This indicates that old samples should be eliminated from the stacks, which will be discussed in the sample selection procedure in Sec.~\ref{sec:hissm}.
\end{remark}
To assist the following analysis, we define a residual signal
\begin{equation}\label{eq:filt}
\xi_d(t) = \dot{d}(t) - \Lambda d(t)
\end{equation}
and the accumulated error
\begin{equation}\label{eq:remt}
\textstyle \xi(\tau_s,t) \!=\! \xi_d(t)+ \sum^{n_s}_{j=1}  \int^{t}_{t_{j}} S(t_j,t) e^{\Lambda (t-\tau)} \xi_d(\tau) \mathrm{d} \tau. 
\end{equation}
Then, the convergence of the estimation error $\tilde{d}(t) \!=\! d(t) - \hat{d}(t)$ is given by the following theorem.

\begin{theorem}\label{th:bfet}
For the dynamic system in (\ref{eq:sysori}) and the disturbance observer in (\ref{eq:obs}), the estimation error $\tilde{d}(t)$ is uniformly ultimately bounded (UUB) by 
\begin{equation}
\textstyle \mathcal{D}(\varrho) \!=\! \left\{ \tilde{d}(t)\! \left| \|\tilde{d}(t) \| \!<\! \frac{\varrho+1}{\omega} \overline{\xi}_t \right. \right\},
\end{equation}
where $\overline{\xi}_t \in \mathbb{R}^+$ is the upper bound of $\xi(\tau_s,t)$, for all $x(0) \in \mathbb{R}^n$, $d(0) \in \mathbb{R}^p$, and $\varrho \in \mathbb{R}^+$, if there exists $\omega \in \mathbb{R}^+$, such that the history stack $S(\tau_s,t)$ satisfies
\begin{equation}\label{eq:th3}
\textstyle S(\tau_s,t) > \frac{\omega I + \Lambda}{\kappa}.
\end{equation}
\end{theorem}

\begin{proof}
See Sec. \ref{sec:proof1}.
\end{proof} 

\begin{remark}\label{rm:th1}
Theorem \ref{th:bfet} shows that the estimator (\ref{eq:obs}) guarantees a UUB property $\|\tilde{d}(t)\| \in \mathcal{D}(\varrho)$ which regulates the precision level of the disturbance estimation. With a determined $\omega$, the bounding scalar $\overline{\xi}_t$ of the accumulated error $\xi(\tau_s,t)$ is the main factor that affects the ultimate error bound $\mathcal{D}(\varrho)$. For (\ref{eq:remt}), there exists $t_j < t'_j <t$ for each sample, such that
\begin{equation*}
\begin{split}
& \textstyle \|\xi(\tau_s,t)\| \leq  \xi_d(t) + \sum^{n_s}_{j=1}  \int^{t}_{t_{j}} \| S(t_j,t) e^{\Lambda (t-\tau)} \xi_d(\tau)\| \mathrm{d} \tau \\
& \textstyle ~~~~= \xi_d(t) + \sum^{n_s}_{j=1} (t-t_j) \|S(t_j,t)  e^{\Lambda (t-t'_j)} \xi_d(t'_j)\| \\
& \textstyle ~~~~\leq \| \xi_d(t)\|\!\left(1+ \sum^{n_s}_{j=1} (t-t_j) \|S(t_j,t) \|\| e^{\Lambda (t- t'_j)}\|\right). 
\end{split}
\end{equation*}
Therefore, the bounding scalar $\overline{\xi}_t$
is determined by both the residual signal $\xi_d(t)$ and the matrix norms $\|S(t_j,t) \|$ and $\| e^{\Lambda (t-t'_j)}\|$. While the former is mainly determined by the property of the disturbance $d(t)$, the latter are affected by the time increment $t-t_j$ which depicts the accumulation of the residual signal $\xi_d(t)$ as time increases. In this sense, the time increment $t-t_j$ should be confined to limit the accumulative error $\xi(\tau_s,t)$, which addresses a similar concern to Remark \ref{rm:boundelt}. This can be achieved by eliminating the old samples from the stacks, which will be considered for the history sample selection procedure in Sec.~\ref{sec:hissm}. Besides, the constant $\kappa$ can also adjust the bound $\overline{\xi}_t$ by releasing the requirement on the stack $S(\tau_s,t)$. With a larger $\kappa$, fewer samples are needed in the history stack $S(\tau_s,t)$ and the bound $\overline{\xi}_t$ can be restricted.
\end{remark}

The proposed CL-based disturbance observer is equivalent to the conventional disturbance observers proposed in~\cite{ciccarella1993luenberger, Chen2000}, where the history stacks only contain the most recent sample, i.e., $\tau_s \!=\! \{t\}$ and $S(\tau_s,t) \!=\! L^{\top}(x(t))L(x(t))$, $\forall \, t \in \mathbb{R}^+$. In this situation, the convergence condition (\ref{eq:th3}) may not hold when the disturbance gain $L(x(t))$ is close to the singularity which reflects the lack of PE according to (\ref{eq:pers}). Compared to the conventional disturbance observers, the CL-based observer can still ensure the convergence condition (\ref{eq:th3}) due to the accumulation of the history stack $S(\tau_s,t)$, even when PE is not satisfied. Besides, compared to the constant $\kappa$, $S(\tau_s,t)$ can adaptively adjust the feedforward gain of the observer to avoid it being overlarge. Note that an overlarge gain may cause instability of the closed-loop dynamics of the observer with finite sampling frequency. This will be further discussed for the discrete-time form in the next subsection.

\subsection{Disturbance Observer in Discrete Time}\label{sec:dedist}

For the application of the method to discrete-time systems, we present the discrete-time form of the disturbance observer (\ref{eq:obs}) considering the finite sampling rate. We study the following discrete-sampled system as a substitution of (\ref{eq:sysreg2}),
\begin{equation}\label{eq:dissys}
 z(k+1) = z(k) + h \gamma \!\left(x(k),u(k) \right) +h B L\!\left(x(k)\right)\!d(k) , 
\end{equation}
where $x(k) \in \mathbb{R}^n$, $z(k) \in \mathbb{R}^r$, $u(k) \in \mathbb{R}^m$ and $d(k) \in \mathbb{R}^p$ are respectively the system states, input and disturbance at sampling instant $k \in \mathbb{N}_{\geq0}$, and $h$ is the sampling period. The discrete-time disturbance observer is formulated as
\begin{equation}\label{eq:disobs}
\hat{d}({k+1}) = \left( e^{h \Lambda} - \kappa h S(\kappa_s,k) \right)\!\hat{d}(k) + \kappa X(\kappa_s,k),
\end{equation}
where $\hat{d}(k) \in \mathbb{R}^p$ is the estimation of $d(k)$, $\kappa_s=\{k_1$, $\cdots$, $k_{n_s}\}$ are the sampling instants of the history data, and 
\begin{equation*}
\textstyle S(\kappa_s,k) = \sum^{n_s}_{j=1} S(k_j,k),~X(\kappa_s,k) = \sum^{n_s}_{j=1} X(k_j,k),
\end{equation*}
are the discrete-time history stacks, where
\begin{subequations}\label{eq:disst}
\begin{equation}\label{eq:diss2}
S(k_j,k) \!=\! e^{h\Lambda(k_j-k)}\! L^{\top}_j L_j e^{h\Lambda(k_j-k)},
\end{equation}
\begin{equation}\label{eq:disr2}
X(k_j,k) \!=  \! e^{h\Lambda(k_j-k)}\! L^{\top}_j \!B^{\top\!} \breve{\zeta}_j,
\end{equation}
\end{subequations}
where $L_j = L(x(k_j))$ and $\breve{\zeta}_j$ is a difference term
\begin{equation}\label{eq:diszeta}
 \breve{\zeta}_j= h\gamma(x(k_{j}),u(k_{j})) + \psi \!\left(x(k_{j}) \right) - \psi \!\left(x(k_{j}+1)\right).
\end{equation}
where $x(k_j)$ and $u(k_j)$ are respectively the system state and input at sampling instant $k_j$.

Compared to the continuous-time observer (\ref{eq:obs}), the discrete-time form (\ref{eq:disobs}) does not need the state derivatives but requires the successor states instead. Therefore, for any sampling instant $k$, the current state $x(k)$ is not stacked since its successor state $x(k+1)$ is not available. Similar to (\ref{eq:remt}), we define the discrete-time residual signal
\begin{equation}\label{eq:filtdis}
\xi_d(k) = d(k+1) - e^{h\Lambda} d(k)
\end{equation}
and the accumulated error
\begin{equation*}
\textstyle \xi(\kappa_s,k) \!=\!\xi_d(k)+ \sum^{n_s}_{j=1}\! \sum^{k-1}_{i=k_j}\! e^{h\Lambda(k_j\!-\!k)}\! L^{\top}_j L_j e^{h\Lambda(k_j\!-\!k)} \xi_d(i).
\end{equation*}
Then, the convergence of the estimation error $\tilde{d}(k) = d(k) - \hat{d}(k)$ is given by the following theorem.
{
\begin{theorem}\label{th:theo5}
For the discrete-time system (\ref{eq:dissys}) and the disturbance observer (\ref{eq:disobs}), the estimation error $\tilde{d}(k)$ is UUB by
\begin{equation}
\textstyle \breve{\mathcal{D}}(\varrho) \!=\! \left\{ \tilde{d}(k) \left|  \| \tilde{d}(k) \| \! < \! \left(\frac{1}{2\omega}\!+\! \sqrt{\frac{1}{4\omega}\!+\!\frac{h}{\omega}} \right)\! (\varrho\!+\!1) \overline{\xi}_k \right. \right\}
\end{equation}
for all $x(0)\in \mathbb{R}^n$, $d(0) \in \mathbb{R}^p$, and $\varrho \in \mathbb{R}^+$, if there exists $\omega \in \mathbb{R}^+$, such that the history stack $S(\kappa_s,k)$ satisfies
\begin{equation}\label{eq:co51}
 S_L < S(\kappa_s,k) < S_U,
\end{equation}
where $\overline{\xi}_k$ is the upper bound of $\xi(\kappa_s,k)$ and

\begin{subequations}\label{eq:S_bound}
\begin{alignat}{2}
\textstyle S_L = \frac{1}{h \kappa}\!\left( e^{h \Lambda} - \left( \frac{1}{2}\!+\! \sqrt{\frac{1}{4}\!-\!h\omega} \right)\!I \right), \\
\textstyle S_U = \frac{1}{h \kappa}\!\left( e^{h \Lambda} - \left( \frac{1}{2}\!-\! \sqrt{\frac{1}{4}\!-\!h\omega} \right)\!I \right)
\end{alignat}
\end{subequations}
are constant matrices prescribing the bounds of $S(\kappa_s,k)$.

\end{theorem}
}
\begin{proof}
See Sec. \ref{sec:proof7}.
\end{proof} 

\begin{remark}
Compared to the continuous-time form (\ref{eq:obs}), the discrete-time observer (\ref{eq:disobs}) renders a more strict convergence condition due to the sampling period $h$, although both forms ensure the estimation errors to be UUB. Besides, the ultimate error bound $\breve{\mathcal{D}}(\varrho)$ is more conservative than $\mathcal{D}(\varrho)$ with an additional multiplier $o_1 (h \omega) \!>\! 1$, $\forall \, h \in \mathbb{R}^+$. In an extreme case, $\breve{\mathcal{D}}(\varrho) \rightarrow \mathcal{D}(\varrho)$ as $h \rightarrow 0$. Note that an overlarge gain $\kappa$ may violate the convergence condition (\ref{eq:co51}) and leads to imprecise estimation results. Thus, the history stack $S(\kappa_s,k)$ can adaptively adjust the gain of the observer to consistently ensure the error convergence.
\end{remark}

\begin{remark}
Note that the convergence condition in~\eqref{eq:co51} is subject to bilateral constraints. If the value of $h \kappa$ is too large, there may exist cases when~\eqref{eq:co51} is not feasible. To avoid this issue, either $h$ or $\kappa$ should be selected as small. 
\end{remark}

The CL-based disturbance observer has been presented in both continuous-time and discrete-time forms. Although $X(\tau_s,t)$ and $X(\kappa_s,k)$ respectively use the state derivative $\dot{x}(\tau_s)$ and the subsequent state $x'(\kappa_s)$, $S(\tau_s,t)$ and $S(\kappa_s,k)$ have the same structure with the correspondence $t=hk$. Also, for both forms, the estimation errors are mainly affected by the accumulated errors which are inevitable for the CL-based disturbance observer since the disturbance is actuated by the unknown extrinsic dynamics. In this sense, the accumulated errors reflect the compromise of the estimation to the imperfect knowledge of the disturbance. Note that the arguments on the continuous-time accumulated error $\xi(\tau_s,t)$ in Remarks \ref{rm:phyt} and \ref{rm:th1} also hold for the discrete-time one $\xi(\kappa_s,k)$. In the next subsection, we will discuss how the accumulated errors can be restricted using a history-sample-selection procedure.

\subsection{History Sample Selection}\label{sec:hissm}

As addressed in Remarks~\ref{rm:boundelt} and~\ref{rm:phyt}, too much data in the history stacks may lead to large accumulated errors. Also, to ensure the convergence of the estimation errors, the history stack $S(\kappa_s, k)$ in \eqref{eq:co51} is limited by both upper and lower bounds. Thus, a procedure is needed to limit the amount of the history data in the stacks to ensure both the static and dynamic performance of the estimation, namely the ultimate estimation accuracy and the convergence of the estimation errors, respectively. Similar procedures used to purge stacks and remove erroneous data are referred to as \textit{history stack management} in previous work on CL~\cite{kersting2015removing}. In this subsection, we propose a history sample selection procedure, shown in Algorithm~\ref{ag:rme}, to resolve this problem. The algorithm is only presented in discrete time for brevity, although it can be adapted to continuous time according to the inverse discretization $t=hk$. It requires the current time $k$, the sampling instants $\kappa_s$, the terms $\breve{\zeta}_j$ and $L_j$ for all $j \in \kappa_s \cup k-1$, and the history stacks $S$ and $X$. The main technical points of the history sample selection procedure are introduced as follows.

\subsubsection{Updating New Sample}

At every run-time instant $k \in \mathbb{N}^+$, we add new data to the history stacks. Line 1 stores the latest instant $k-1$ to the queue $\kappa_s$. Line 2 and Line 3 use incremental approaches to update the history stacks $S$ and $X$.

\subsubsection{Purging Old Samples}

After collecting new data, we purge the old data from the history stacks. The principle is that we always start purging from the oldest sampling instant. The objective of the purging is to ensure that condition \eqref{eq:co51} holds. Also, the stack $S(\kappa_s,k)$ should be as close to $S_L$ as possible to guarantee a small error bound. As shown in the \textit{for}-loop between Line 4 and Line 12, we purge the old samples one by one, until $S(\kappa_s,k) > S_L$ is just satisfied and is about to be violated. The bounds $S_L$ and $S_U$ are calculated using \eqref{eq:S_bound} with a feasible $\omega$. The symbol END refers to the ending index of the queue $\kappa_s$.

\begin{algorithm}[htbp]
\caption{History Sample Selection Procedure}
\label{ag:rme}
\begin{algorithmic}[1]
\renewcommand{\algorithmicrequire}{\textbf{Input:}}
\renewcommand{\algorithmicensure}{\textbf{Output:}}
\REQUIRE $k$, $\kappa_s$, $\zeta_j$, $L_j$, for all $j \in \kappa_s \cup k-1$, $S$, $X$
\ENSURE $\kappa_s$, $S$, $X$
        \STATE $\kappa_s = \kappa_s \cup k-1$
        \STATE $S = e^{-h\Lambda} \!\left( S \!+\! L^{\top}_{k-1}L_{k-1} \right)\! e^{-h\Lambda}$
        \STATE $X = e^{-h\Lambda} \!\left( X \!+\! L^{\top}_{k-1}\zeta_{k-1}\right)$
        \FOR {$j= 1$ \TO END} 
        \STATE {$S' = S - e^{h\Lambda(k_j-k)} L^{\top}_jL_j e^{h\Lambda(k_j-k)}$}
            \IF  {$S' > S_L$ \OR $S' > S_U$}
                \STATE {$S = S'$}
                \STATE {$X = X - e^{h\Lambda(j-k)} L^{\top}_j\zeta_j$}
            \ELSE
                \STATE {\textbf{break}}
            \ENDIF
        \ENDFOR
        \STATE $\kappa_s = \kappa_s(j:\mathrm{END})$
\end{algorithmic} 
\end{algorithm}


\begin{remark}
Algorithm \ref{ag:rme} does not affect the convergence of the estimation error, although it renders a non-trivial sampling process for the history stacks. The reason is that the procedure only adapts the feed-forward gain of the closed-loop of the estimation but does not interfere with its stability.
\end{remark}

The result of Algorithm \ref{ag:rme} is that only the newest samples are kept in the history stacks and the accumulated errors are restricted to the lowest possible level. When the disturbance gain approaches the singularity, more samples will be stacked to ensure the convergence of the estimation error, which renders a PE-free method. Otherwise, redundant samples will be eliminated and the observer behaves closely to a conventional PE-based disturbance observer. Thus, the CL-based disturbance observer ensures precise estimation when PE is not ensured while maintaining a low error bound when PE is satisfied. This indicates its advantage compared to the conventional observers in terms of both flexibility and adaptability.


\section{Simulation Study: Epidemic Control}\label{sec:simu}

In this section, we evaluate the proposed CL-based disturbance observer in an epidemic-control case that simulates the spread of the epidemic over a social network. The networked epidemic model is widely used to characterize the epidemic spreading process and to predict the contiguous states where the population is not well-mixed. Sec.~\ref{sec:preC} addressed that the model may lose the PE condition when the network is close to zero infection probabilities. Therefore, it is an ideal model to validate the advantage of our proposed method over conventional disturbance observers. Note that the epidemic model is just a baseline model that can best reflect the advantage of the proposed methods. In theory, the proposed method can be applied to any PE-free dynamic systems, including circuit or grid networks~\cite{uchida2019incentivizing}.

The study applies the proposed observer to the estimation of the infection rates of the epidemic to improve the control scheme. A comparative study is conducted to show the superior precision of the proposed observer over the conventional disturbance observer when the system fails to ensure PE. The simulation is conducted in MATLAB R2020b using the first-order Euler solver. The hardware to run the simulation is a Thinkpad Laptop without graphic cards. The simulation program and the dataset can be referred to in our GitHub repository~\cite{github}.

\subsection{The Networked Epidemic Model}\label{sec:chal}

In this study, we use the discrete-time susceptible-infected-susceptible (SIS) model~\cite{liu2019robust, liu2021optimal} to formulate the spread of epidemics over a social network. It considers a weighted digraph $\mathcal{G} \!=\! (\mathcal{V},\mathcal{E},W)$ with $n \in \mathbb{N}^+$ nodes, where $\mathcal{V} \!=\! \{1,2,\ldots, n\}$ and $\mathcal{E} \subseteq \mathcal{V} \!\times\! \mathcal{V}$ are respectively the set of vertices and the set of edges. $W \!=\! [w_{ij}] \!\in\! \mathbb{R}^{n \times n}$ is the adjacency matrix of $\mathcal{G}$. Here, we only consider a graph $\mathcal{G}$ with no self-loops, i.e., $w_{ii} \!=\! 0$, $\forall \, i \in \mathcal{V}$. We also confine ourselves that $w_{ij} > 0$, $\forall \, i,j \in \mathcal{V}$, if there exists an edge from $j$ to $i$. Then, the discrete-time dynamic model of the SIS model is~\cite{liu2016node}
\begin{equation}\label{eq:infepi1}
\textstyle x_i(k\!+\!1) \!=\! h\! \left(1\!-\!x_i(k)\right) \sum^n_{j=1}\! w_{ij} d_j x_j(k)\! +\! \left(1\!-\!h \delta_i \right)\!x_i(k),
\end{equation}
$i \in \mathcal{V}$, where for each note $i$, $x_i(k)$ denotes its infection probability at time step $k \!\in\! \mathbb{N}_{\geq 0}$, $d_i \in \mathbb{R}_{\geq 0}$ is the proactive infection rate, $\delta_i \in \mathbb{R}_{\geq 0}$ is the passive curing rate at instant $k$, and $h \in \mathbb{R}^+$ is the sampling period. Under appropriate assumptions~\cite{pare2018analysis}, the infection probability $x_i(k)$, for all $k \in \mathbb{N}_{\geq0}$ and $i \in \mathcal{V}$, is well defined, i.e., given any initial condition $x_i(0) \in \left[\,0,\,1\, \right]$, the system state is confined by $x_i(k) \in \left[\,0,\,1\, \right]$. Note that this model~\eqref{eq:infepi1} is the discretization of the continuous-time model in \eqref{eq:contep}.

In practice, the graph $\mathcal{G}$ of the SIS model represents the social network of the people in a community. Each node $i \in \mathcal{V}$ is recognized as an individual. The edge set $\mathcal{E}$ and the adjacency matrix $W$ denote the consistent contacts among the nodes. The system state $x_i(k)$ represents how likely an individual is infected in the statistical sense. The infection rate $d_i$ measures how susceptible an individual is towards the epidemic and the curing rate $\delta_i$ describes how easily one recovers from the epidemic. High infection rates may bring up the infection probabilities while high curing rates perform otherwise. In general, people with inferior immunity usually correspond to high infection rates and those receiving positive treatments tend to have higher curing rates. Also, the infection rates may slowly change due to seasons, foods, or health conditions. On the contrary, the curing rates can be improved by manual intervention like active medical treatments. The essential objective of epidemic control is to design a control scheme for the curing rate, such that the infection probabilities  are driven to zeros with the existence of the infection rates.

\subsection{Experimental Configuration}

\begin{figure}[htbp]
 \centering 
 \includegraphics[width=0.25\textwidth]{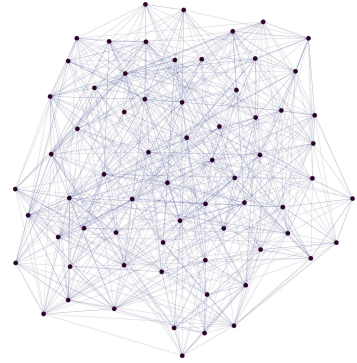}
 \caption{The visualization of the $67$-node simulated social network. An edge between two nodes indicates the social connection between two people.}
 \label{fig:network} 
\end{figure}

\begin{figure}[htbp]
 \centering 
 \includegraphics[width=0.45\textwidth]{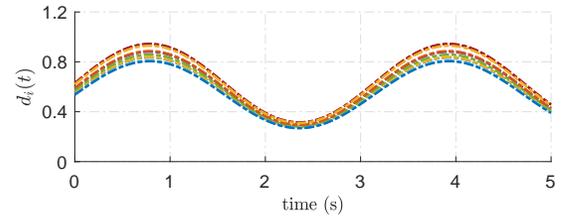}
 \caption{The true infection rates of the nodes in the network with timing discretization $t = hk$. Each line represents the infection rate of a node. Note that we only show nodes $i=1,8,15,\cdots,64$, for brevity.}
 \label{fig:dist} 
\end{figure}

\begin{figure}[htbp]
 \centering 
 \includegraphics[width=0.45\textwidth]{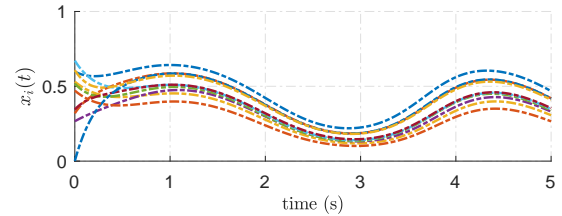}
 \caption{The infection probabilities of the network with timing discretization $t = hk$. Each line represents the infection probability of a node. Note that we only show nodes $i=1,8,15,\cdots,64$, for brevity.}
 \label{fig:x} 
\end{figure}

In this experiment, we use a graph containing $n=67$ nodes to represent the social connection of the residences in a community. The connection of the graph is visualized in Fig.~\ref{fig:network}. It is noticed that this graph has high complexity due to both its large scale and the strong connection between its nodes, which is sufficient to validate the efficacy of the proposed method. We use the networked epidemic model in~\eqref{eq:infepi1} to simulate the spreading of the disease. The parameters of the model, namely the adjacency matrix $W$, the initial infection probabilities $x_i(0)$, the ground truth of disturbance $d_i$, and the baseline curing rate $\bar{\delta}_i$, $i=1,2,\cdots,n$, can be found in our online repository~\cite{github} and are not enumerated due to the large scales.  We assume the infection rates to be time-variant and sinusoidal as shown in Fig.~\ref{fig:dist} to simulate their changes as time increases. Also, the initial infection probabilities are sampled from a uniform distribution $x_i(0) \!\sim\! \mathsf{U}(0,1)$, $i=1,2,\cdots,n$, to simulate general cases. The simulation duration is $T=5$ with a discrete sampling rate $h\!=\!10^{-4}$. The infection probabilities of the nodes under the influence of the given infection rates are shown in Fig.~\ref{fig:x}. It is noticed that the infection probabilities are close to zero values around the time instant $t\!=\!3$. This indicates that the disturbance gain $(I- \mathrm{diag}(x_t))W \mathrm{diag}(x_t)$ of the epidemic model~\eqref{eq:contep} is close to singularity around this timing point and the system loses PE.

\subsection{Comparison with Conventional Method}\label{sec:simuconf}

\begin{figure}[htbp]
	\centering
	\subfigure[The disturbance estimation errors of the CL-based observer with timing discretization $t=hk$. Each line shows the estimation error of a node. Note that we only show nodes $i=1,8,15,\cdots,64$, for brevity.]{\label{fig:vldrsl1}\includegraphics[width=0.45\textwidth]{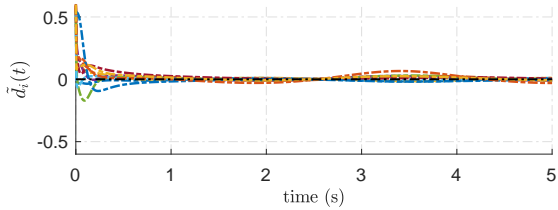}}		
    \subfigure[The depth of the history stacks of the CL-based disturbance observer $n_s$ at each instant.]{\label{fig:vldrsl3}\includegraphics[width=0.45\textwidth]{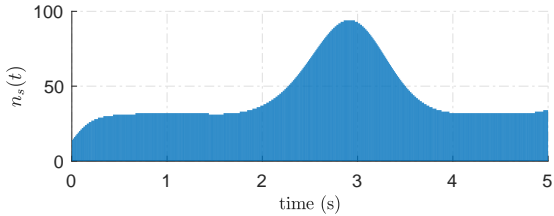}}	
	\subfigure[The disturbance estimation errors of the conventional observer with timing discretization $t=hk$. Each line shows the estimation error of a node. Note that we only show nodes $i=1,8,15,\cdots,64$, for brevity.]{\label{fig:vldrsl2}\includegraphics[width=0.45\textwidth]{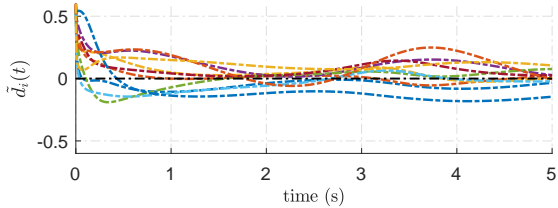}}		
			
	\caption{The performance of the CL-based and the conventional observers.}
	\label{fig:vldrsl}
\end{figure}

In this subsection, we use the proposed CL-based disturbance observer \eqref{eq:disobs} to estimate the time-variant infection rates shown in Fig.~\ref{fig:dist}. 
The parameter of the CL-based observer is set as $\kappa=100$. The value of parameter $\Lambda$ can be referred to in~\cite{github} and is not listed here due to limited space. The initial conditions of the observer are set to zero values. Algorithm~\ref{ag:rme} is used for history data selection with the parameter $\omega=5$. To address the advantage of the proposed CL-based observer, we also use a classical and widely used disturbance observer in~\cite{Chen2000} to perform a comparison study. As addressed in Sec.~\ref{sec:eccd}, this observer is equivalent to a one-depth-stack CL disturbance observer. Therefore, we design the conventional observer also in a CL-based form but only store the latest sample $x(k-1)$ in the history stacks. For a fair comparison, the conventional observer has the same parameters $\kappa$ and $\Lambda$ as the proposed one. History data selection is not needed for the conventional observer since its history stacks only have one single sample.
The disturbance estimation errors of the CL-based observer and the conventional observer are respectively shown in Fig.~\ref{fig:vldrsl1} and Fig.~\ref{fig:vldrsl2}. The depth of the history stacks of the CL-based observer is also shown in Fig.~\ref{fig:vldrsl3}.

The comparison study indicates that the conventional observer produces larger estimation errors than the CL-based observer. The errors are especially large when the infection probabilities are close zeros and the system loses PE, from $t=3$ to $t=4$. The reason is that PE is not ensured during this time period and the constant gain $\kappa$ is not sufficient to guarantee the convergence of the estimation error. Nevertheless, the CL-based observer still provides precise estimation results even with a slight deviation. This is because that the stacked history samples can provide history data of the system rather than the most recent state. Therefore, we can infer that the history stacks strengthen the convergence condition (\ref{eq:co51}). Fig. \ref{fig:vldrsl3} shows that more samples are stacked when the system is close to the singular states where PE is lost. Specifically, the largest amount appears around $t=3$s when the infection probabilities are closest to zeros (see Fig.~\ref{fig:x}. This indicates that the application of CL is the main reason to ensure the estimation precision of the proposed observer when PE is lost.

\subsection{Disturbance Compensation Control}

\begin{figure}[htbp]
 \centering 
 \includegraphics[width=0.45\textwidth]{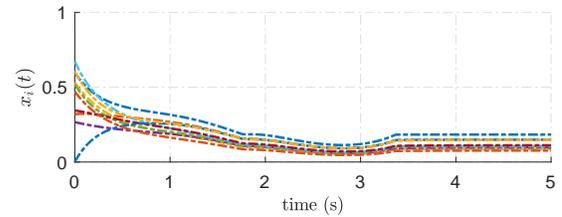}
 \caption{The infection probabilities under the compensation control with timing discretization $t=hk$. Each line shows the infection probability of a node. Note that we only show nodes $i=1,8,15,\cdots,64$, for brevity.}
 \label{fig:sin2} 
\end{figure}

Since the proposed CL-based disturbance observer provides accurate disturbance estimation even when the system does not satisfy PE, it is promising to be used to improve the performance of epidemic control. To verify this, we design a feedforward control law to compensate for the time-variant infection rates during epidemic control. In this case, the curing rate for $i=1,2,\cdots, n$ reads
\begin{equation}\label{eq:ff}
\textstyle \delta_i(k) = \bar{\delta}_i + \left(1\!-\!x_i(k)\right) \sum^n_{j=1}\! w_{ij} \hat{d}_j(k) x_j(k) /x_i(k).
\end{equation}
where $\bar{\delta}_i$ is the baseline curing rate used to generate the infection probabilities in Fig.~\ref{fig:x}. Therefore, based on the baseline curing rates, the control scheme \eqref{eq:ff} has an additional term to compensate for the effects of the infection rates using the estimation. 
The resulting controlled infection probabilities are illustrated in Fig. \ref{fig:sin2}. Compared to Fig.~\ref{fig:x}, it is noticed that Fig.~\ref{fig:sin2} presents a decent control performance since its infection probabilities are consistently retained close to zero values, which means that the epidemics are well controlled. In contrast, Fig.~\ref{fig:x} shows an inferior control result, where the infection probabilities rebound after $t=3$. This simulation study indicates that the epidemic control performance can be greatly improved by estimating the infection rates in run-time and compensating for it in the control input. This use case verifies the potential of the proposed CL-based observer in solving practical problems.


From the simulation results in this section, we notice that the proposed CL-based disturbance observer presents higher estimation precision than the conventional method when PE is not ensured. It also helps to improve the performance of a nominal control scheme. Besides, the application of the proposed method to epidemic control indicates its value in solving practical problems. From this perspective, the target of this paper, to propose a high-precision disturbance observer for a PE-free system, is achieved.

\section{CONCLUSION}\label{sec:concld}

We propose a CL-based disturbance observer to resolve the inferior precision issue for the conventional observer in a PE-free situation. We prove the convergence of the observer using a Lyapunov method and obtain a convergence condition as the substitution of PE. During the application of CL, we notice that large estimation errors may be caused by accumulated errors. To restrict the accumulated errors, we present a history-sample-selection procedure to timely eliminate the old samples. The proposed method serves as a disturbance observer with an adaptive feedforward gain which ensures precise estimation results even when PE is not guaranteed. We use an epidemic control study to show the advantage of the proposed observer over the conventional method and how it can benefit an epidemic control scheme, although it has the potential to be applied to circuit and grid network systems. Therefore, we have addressed both the theoretical feasibility and the implementational effectiveness of the CL-based disturbance observer. It is worth mentioning that the main advantage of the CL-based disturbance observer is the higher disturbance estimation precision in PE-free cases. If PE is satisfied, its performance is very similar to a conventional disturbance observer, since the history stacks do not need to store much history data. In future work, we will investigate its possible applications to a wider range of systems, such as the collision-force estimation of mobile robots with singular configurations.


\section*{Appendix: Proofs}
In the appendix, we provide the proof for the theorems proposed in this article.



\subsection{Proof of Theorem \ref{th:bfet}}\label{sec:proof1}
Subtrating (\ref{eq:obs}) from (\ref{eq:filt}), we obtain the error dynamics
\begin{equation}\label{eq:errdy}
\dot{\tilde{d}}(t) = \Lambda \tilde{d}(t) + \kappa S(\tau_s,t) \hat{d}(t)  - \kappa X(\tau_s,t) + \xi_d(t).
\end{equation}
Note that (\ref{eq:filt}) is a linear model, i.e., for any $t_j \in \mathbb{R}^+$ and $t_j < t$, we have $d(t) = e^{\Lambda (t-t_{j})} d(t_{j}) + \int^{t}_{t_{j}} e^{\Lambda(t-\tau)}\xi_d(\tau) \mathrm{d}\tau$. Considering the non-singularity of $e^{\Lambda (t-t_{j})}$, it leads to
\begin{equation}\label{eq:dmsol2}
\textstyle d(t_{j}) = e^{\Lambda(t_{j}-t)} d(t) \!-\! \int^{t}_{t_{j}} e^{\Lambda(t_j-\tau)} \xi_d(\tau) \mathrm{d} \tau.
\end{equation}
Also, substituting the system dynamics (\ref{eq:sysreg}) to (\ref{eq:zeta}), we obtain $\zeta_j = BL_jd(t_{j})$, $j=1,\cdots,n_s$, which leads the history stack $X(\tau_s,t)$ in (\ref{eq:r}) to
\begin{equation}\label{eq:chi2}
\textstyle X(\tau_s,t)= \sum^{n_s}_{j=1} e^{\Lambda (t_{j}- t)} L^{\top}_j B^{\top\!}\!B L_j d(t_{j}).
\end{equation}
Meanwhile, from (\ref{eq:syspar2}), it is straightforward to infer $B^{\!\top\!}B=I$. Therefore, substituting (\ref{eq:dmsol2}) to (\ref{eq:chi2}), we obtain
\begin{equation*}
\begin{split}
X(\tau_s,t)\!=& \textstyle \sum^{n_s}_{j=1} e^{\Lambda (t_{j}- t)} L^{\top}_j L_j e^{\Lambda(t_{j}-t)} d(t) \\
&\textstyle - \sum^{n_s}_{j=1} \int^t_{t_j} e^{\Lambda (t_{j}- t)} L^{\top}_j L_j e^{\Lambda(t_j-\tau)} \xi_d(\tau) \mathrm{d} \tau\\
=& \textstyle \, S(\tau_s,t) d(t) - \xi(\tau_s,t) + \xi_d(t).
\end{split}
\end{equation*}
Substituting it to (\ref{eq:errdy}), we obtain
\begin{equation}\label{eq:errdy1}
\dot{\tilde{d}}(t) = \, \left( \Lambda- \kappa  S(\tau_s,t) \right)\! \tilde{d}(t) \! +  \xi(\tau_s,t) .
\end{equation}
We define the Lyapunov function as $V(t) = \frac{1}{2} \tilde{d}^{\top}\!(t) \tilde{d}(t)$. Substituting (\ref{eq:errdy1}) to its derivative $\dot{V}(t) = \tilde{d}^{\!\top}\!(t)\dot{\tilde{d}}(t)$, we obtain
\begin{equation*}
 \dot{V}(t) = \, \tilde{d}^{\top\!}\!(t) \left( \Lambda - \kappa S(\tau_s,t) \right) \tilde{d}(t)
 + \tilde{d}^{\top\!}\!(t)\xi(\tau_s,t).
\end{equation*}
Substituting the condition (\ref{eq:th3}) to it, we obtain
\begin{equation*}
 \dot{V}(t) < - \omega \tilde{d}^{\top\!}\!(t) \tilde{d}(t) + \|\tilde{d}(t) \| \left\| \xi(\tau_s,t) \right\|.
\end{equation*}
Since $\|\tilde{d}(t)\| = \sqrt{2V(t)}$ and $\| \xi(\tau_s,t) \| \leq \overline{\xi}_t$, we have
\begin{equation}\label{eq:coro31}
\dot{V}(t) \!< \!- 2 \omega V(t) \!+\!  \overline{\xi}_t\! \sqrt{2V(t)}
\!=\! \sqrt{2V(t)} \!\left( \overline{\xi}_t  \!-\! \omega \! \sqrt{2V(t)}\right).
\end{equation}
Let $\overline{\mathcal{D}}(\varrho)$ be the supplementary set of $\mathcal{D}(\varrho)$. Thus, $\forall  \tilde{d}(t) \in \overline{\mathcal{D}}(\varrho)$, we have $\sqrt{2V(t)} = \| \tilde{d}(t) \| \geq \frac{\varrho + 1}{\omega} \overline{\xi}_t$ which leads (\ref{eq:coro31}) to
\begin{equation}\label{eq:vdotcom}
\dot{V}(t) \!<\! - \varrho \overline{\xi}_t \sqrt{2V(t)} ,
\end{equation}
which ensures $\dot{V}(t) < 0$, $\forall \, \tilde{d}(t) \in \overline{\mathcal{D}}(\varrho)$. Therefore, $\tilde{d}(t)$ ultimately converges to the bounded set $\mathcal{D}(\varrho)$, which uniformly holds for all $x(0) \in \mathbb{R}^n$, $d(0) \in \mathbb{R}^p$, and $\varrho \in \mathbb{R}^+$.


\subsection{Proof of Theorem \ref{th:theo5}}\label{sec:proof7}

Substracting (\ref{eq:disobs}) from (\ref{eq:filtdis}), we obtain the error dynamics
\begin{equation*}
\tilde{d}(k+1) = e^{h \Lambda} \tilde{d}(k) +h \kappa  S(\kappa_s,k) \tilde{d}(k) -\kappa X(\kappa_s,k) + h  \xi_d(k).
\end{equation*}
From (\ref{eq:filtdis}), for any $k_{j} \in \mathbb{N}^+$, $k_{j} < k$, we obtain
\begin{equation*}
\textstyle d(k) = e^{h \Lambda(k-k_j)} d(k_{j}) + \sum^{k-1}_{i=k_j} e^{h \Lambda(k-i-1)}.
\end{equation*}
Considering the non-singularity of $e^{h \Lambda(k-k_j)}$, it leads to 
\begin{equation}
\begin{split}
\textstyle d(k_{j}) = \, e^{h \Lambda(k_j-k)} d(k) - \sum^{k-1}_{i=k_{j}} e^{h \Lambda(k_j-i-1)} \xi_d(i).
\end{split}
\end{equation}
Substituting the discrete-time system (\ref{eq:dissys}) to (\ref{eq:diszeta}), we have
\begin{equation*}
\breve{\zeta}_j = hBL_jd(k_j).
\end{equation*}
Thus, the history stack $X(\kappa_s,k)$ in (\ref{eq:disr2}) leads to
\begin{equation*}
\begin{split}
X(\kappa_s,k)& = \textstyle h \sum^{n_s}_{j=1}\! e^{h \Lambda(k_j-k)}\! L^{\top}_jL_j e^{h \Lambda(k_j-k)} d(k) \\
-& \textstyle h \sum^{n_s}_{j=1} \sum^{k-1}_{i=k_j} \! e^{h \Lambda(k_j-k)}\! L^{\top}_jL_je^{h\Lambda(k_j-k)} \xi_d(i)\\
& \textstyle = hS(\kappa_s,k)d(k) - h\xi(\kappa_s,k) + h \xi_d(k).
\end{split}
\end{equation*}
We define a Lyapunov function
\begin{equation}\label{eq:dis_ly}
\textstyle V(k) = \frac{1}{2} \tilde{d}^{\top\!}(k) \tilde{d}(k).
\end{equation}
Its time increment $\Delta V(k) = V(k+1) - V(k)$ reads
 \begin{equation}\label{eq:dislyadis}
\Delta V(k) = \tilde{d}^{\top\!}(k) \Delta \tilde{d}(k) + \frac{1}{2} \| \Delta \tilde{d}(k)\|^2,
 \end{equation}
where 
\begin{equation*}
\begin{split}
\Delta \tilde{d}(k)\,& = \tilde{d}(k+1) - \tilde{d}(k) \\
&= (\overline{e^{h \Lambda}}- h \kappa S(\kappa_s,k)) \tilde{d}(k) + h \xi(\kappa_s,k),
\end{split}
\end{equation*}
where $\overline{e^{h \Lambda}} = e^{h\Lambda}-I$. Since we have 
\begin{subequations}\label{eq:interV}
\begin{equation}
\tilde{d}^{\top\!}(k) \Delta \tilde{d}(k)\!=\! \tilde{d}^{\top\!}(k)\! \left( \overline{e^{h \Lambda}} \!-\!h \kappa S(\kappa_s,k) \right)\! \tilde{d}(k) +h \tilde{d}^{\top\!}(k)  \xi,
\end{equation}
\begin{equation}
\frac{1}{2} \| \Delta \tilde{d}(k)\|^2 \!\leq\! \tilde{d}^{\top\!\!}(k)\! \left( \overline{e^{h \Lambda}} \!-\!h \kappa S(\kappa_s,k) \right)^{\!2}\! \tilde{d}(k) + h^2 \|   \xi\|^2,
\end{equation}	 
\end{subequations}
and the condition (\ref{eq:co51}) leads to
\begin{equation}\label{eq:condV}
\left( \overline{e^{h \Lambda}} -h \kappa S(\kappa_s,k) \right) + \left( \overline{e^{h \Lambda}} -h \kappa S(\kappa_s,k) \right)^2 < - h \omega I,
\end{equation}
substituting (\ref{eq:interV}) and (\ref{eq:condV}) to (\ref{eq:dislyadis}) leads to	
\begin{equation*}
	\Delta V(k) < - h\omega \|\tilde{d}(k)\|^2 +h \|\tilde{d}(k)\| \|\xi\| + h^2 \|\xi\|^2.
\end{equation*}
Considering $\|\tilde{d}(k)\| = \sqrt{2V(k)}$ and $\| \xi \| \leq \overline{\xi}_k$, we have
\begin{equation}
	\Delta V(k)< -2h\omega V(k) + h \overline{\xi}_k \sqrt{2V(k)}  + h^2 \overline{\xi}_k ^2 .
\end{equation}
Using \eqref{eq:dis_ly}, it leads to
\begin{equation}\label{eq:cr4-3}
	\Delta V(k)< -h\omega \|\tilde{d}(k)\|^2 + h \overline{\xi}_k \|\tilde{d}(k)\|  + h^2 \overline{\xi}_k ^2 .
\end{equation}
Letting $\overline{\breve{\mathcal{D}}}(\varrho)$ be the complementary set of $\breve{\mathcal{D}}(\varrho)$, we have 
\begin{equation}\label{eq:dissolV}
	\Delta V(k)< -h\omega \|\tilde{d}(k)\|^2 + h \overline{\xi}_k \|\tilde{d}(k)\|  + h^2 \overline{\xi}_k ^2 < 0,
\end{equation}
for all $\tilde{d}(k) \in \breve{\mathcal{D}}(\varrho)$, $\varrho \in \mathbb{R}^+$
Note that (\ref{eq:dissolV}) uniformly holds for all $V(0) \in \mathbb{R}^+$, which indicates that $\tilde{d}(k)$ is uniformly ultimately bounded by $\breve{\mathcal{D}}(\varrho)$.



\ifCLASSOPTIONcaptionsoff
 \newpage
\fi



%


\balance

\end{document}